\documentclass[11pt]{article}
\usepackage{graphicx}
\usepackage{amsmath}
\usepackage{amssymb}
\usepackage{amsxtra}

\title{Balancing baryon and asymmetric dark matter excess}
\author{V. A. Beylin$^{1,3}$, M. Yu. Khlopov$^{1,2,3}$, D. O. Sopin$^2$\\$^1$ Institute of Physics,
Southern Federal University\\
Stachki 194
Rostov on Don 344090, Russia\\ $^2$ National Research Nuclear University MEPhI 115409 Moscow, Russia\\$^3$ Virtual Institute of Astroparticle physics 75018 Paris, France\\
khlopov@apc.univ-paris7.fr}

\begin{document}
\maketitle

\begin{abstract}
Effect of the electroweak non-conservation of the baryon number could be a key ingredient to explain the ratio of dark and baryonic densities. If dark matter is explained by dark atoms, in which stable $-2n$ charged particles are bound with $n$ nuclei of primordial helium, and this multiple charged particles possess SU(2) electroweak charges, the excess of $-2n$ charged particles over their antiparticles can be related to baryon excess by sphaleron transitions. It provides relationship between the density of asymmetric dark atom dark matter and baryon asymmetry,  The cosmological consequences of sphaleron transitions were considered for the minimal walking technicolor (WTC) model, which provides composite Higgs boson solution for the problem of Higgs boson mass divergence in the Standard model. The realisation of multi-component dark atom scenario is possible because the electric charges of new fermions are not fixed and several types of stable multiple charged states are possible. In particular cases the upper limits for the  masses of techniparticles could be found, at which dark atom interpretation of dark matter is possible. These limits challenge search for multiple charged stable particles at the LHC. 
\end{abstract}

\section{Introduction}\label{s:intro}

The modern cosmological paradigm inevitably involves the dark matter but it's origin and dynamics should be described by physics beyond the Standard Model. In this paper the consequences of the minimal walking technicolor model (WTC) are considered \cite{WTC_Sannino_2005,WTC_Hong_2004,WTC_Tuominen_2005,WTC_Dietrich_2006}. 

The existing of four new fermions is assumed, namely: heavy  techniquarks U, D have a standard electroweak charges and the new one --- technicolor charge. These particles could be observed in the form of technibaryons ($UU$, $UD$, $DD$) which arise as a Goldstone bosons from the global symmetry breaking, $SU(4)\rightarrow SO(4)$. In the model, the Higgs boson becomes a composite particle $\frac{1}{\sqrt{2}}\left(U\bar{U}+D\bar{D}\right)$ and technileptons, N and E, are introduced to eliminate anomalies.

Importantly, an (arbitrary) electric charge of techniparticles is not fixed and can be specified by $y$-parameter: ($y+1$, $y$, $y-1$, $\frac{1}{2}(-3y+1)$, $\frac{1}{2}(-3y-1))$ for $(UU,\,UD,\,DD,\,N,\,E)$ states, correspondingly. 

In minimal WTC model the bound state $X^{-2n}(^4He^{+2})_n$, where the stable heavy core, $X$, is $(UU)^{-2n}$ or $(E/N)^{-2n}$, could be considered as a dark matter particle. Such "dark atom"\,, called also X-helium, fulfill conditions of no-go theorem which is formulated in \cite{nongo_Khlopov, nongo_Belotsky_2006}.

It is assumed that the required excess of techniparticles is generated during nonperturbative electroweak processes violating the laws of baryon and lepton numbers conservation. Such processes are generated by the topology of the $SU(2)$ group and should actively occur in the early Universe. In the literature, these stable classic solutions are called sphaleron transitions \cite{Sph_Manton, Sph_Klinkhamer}.

The cosmological consequences of this model have already been partially considered in \cite{Kouv1, Kouv2, Kouv3}. Authors used the thermodynamic approach to balance baryon and dark matter excesses for the case $y=1$, however, the general case has not been studied.

This paper is organized as follows. In section \ref{s:eqa} equations for chemical potentials are written out. The solutions for temperatures before and after the electroweak phase transition (EWPT) are considered in sections \ref{s:before} and \ref{s:after}, correspondingly. Some discussion of the results is presented in Conclusions \ref{s:concl}.

\section{Equations}\label{s:eqa}

The thermodynamic approach for the analysis of sphaleron configurations was developed in \cite{Kouv1, Kouv2, Kouv3, HarvTurn}. In this paper, chemical potentials are introduced in a similar way: $\mu_{iR/L}$, where "$i$"  is a flavor of particle and $R/L$ is the chirality. So the conditions of weak interactions can be written as following: 
\begin{align}
    \mu_{iR}=\mu_{iL}\pm\mu_0,
    \\
    \mu_{i}=\mu_{j}+\mu_W,
\end{align}
where "i"- and "j"-type fermions form an electroweak doublet.

The standard baryon number and lepton number densities are 
\begin{align}
&\begin{aligned}
B &= \cfrac{1}{3}\cdot 3\cdot (2+\sigma_{t})(\mu_{uL} + \mu_{uR}) + \cfrac{1}{3}\cdot 3 \cdot 3\cdot (\mu_{dL} + \mu_{dR})=
\\
& = (10 + 2\sigma_{t})\mu_{uL} + 6\mu_{W},
\label{B}
\end{aligned}
\\
&\begin{aligned}
L &= \sum_i (\mu_{\nu_iL} + \mu_{\nu_iR} + \mu_{iL} + \mu_{iR})=
\\
&=  4\mu + 6\mu_{W},
\label{L}
\end{aligned}
\end{align}
where the weight functions for a massive particles are used:
\begin{equation}
\label{eq: sigma-function def}
   \sigma(z)=
    \begin{cases}
    \frac{6}{4\pi^2}\int_{0}^{\infty}{dx ~x^2 \cosh^{-2}{\left(\frac{1}{2}\sqrt{x^2+z^2   }\right)}}, \; \text{for fermions;} \\
    \frac{6}{4\pi^2}\int_{0}^{\infty}{dx ~x^2 \sinh^{-2}{\left(\frac{1}{2}\sqrt{x^2+z^2   }\right)}}, \; \text{for bosons.}
    \end{cases}
\end{equation}

The number densities for technibaryons and technileptons are similar to the ones, defined in \cite{Kouv3}:
\begin{align}
&\begin{aligned}
TB=\cfrac{2}{3}(\sigma_{UU}\mu_{UU}+\sigma_{UD}\mu_{UD}+\sigma_{DD}\mu_{DD}),
\label{TB}
\end{aligned}
\\
&\begin{aligned}
TL=\sigma_E(\mu_{EL}+\mu_{ER})+\sigma_N(\mu_{NL}+\mu_{NR}),
\label{TL}
\end{aligned}
\end{align}

The solution of this system of equations is an observable ratio of densities which strongly depends on the sphaleron's freezing out temperature $T_*\sim 200 \,\mbox{GeV}$. If it is higher then the temperature of EWPT $T_{EWPT}$, the condition of isospin neutrality could be used:
\begin{align}
&\begin{aligned}
0=&\cfrac{1}{2}\cdot3\cdot3\cdot(\mu_{uL}-\mu_{dL})+\cfrac{1}{2}\cdot3\cdot3\cdot(\mu_{iL}-\mu_{eL})+
\\
&+\sigma_{UU}\mu_{UU}-\sigma_{DD}\mu_{DD}+\cfrac{1}{2}\sigma_N\mu_{NL}-\cfrac{1}{2}\sigma_E\mu_{EL}-4\mu_W-\mu_W.
\end{aligned}
\label{I32}
\end{align}

The equation of sphaleron transition is also similar to the one, defined in \cite{Kouv3}:
\begin{equation}
    3(\mu_{uL}+2\mu_{dL})+\mu+\cfrac{1}{2}\mu_{UU}+\mu_{DD}+\mu_{NL}=0.
\end{equation}
However, the electrical neutrality condition should be slightly modified:
\begin{align}
&\begin{aligned}
0=&\cfrac{2}{3}\cdot3\cdot3(\mu_{uL}+\mu_{uR})-\cfrac{1}{3}\cdot3\cdot3(\mu_{dL}+\mu_{dR})-\cfrac{1}{3}(\mu_{eL}+\mu_{eR})+
\\
&+(y+1)\sigma_{UU}\mu_{UU}+y\sigma_{UD}\mu_{UD}+(y-1)\sigma_{DD}\mu_{DD}+
\\
&+\cfrac{-3y+1}{2}\sigma_{N}(\mu_{NL}+\mu_{NR})+\cfrac{-3y-1}{2}\sigma_{E}(\mu_{EL}+\mu_{ER})-4\mu_W-2\mu_m,
\end{aligned}
\end{align}
where the charges of the techniparticles are parameterised by $y$.

The last condition describes the composite nature of Higgs boson and can be used in both cases, because the chemical potentials of particle and antiparticle differ only by sign:
\begin{equation}
    \mu_0=0.
    \label{mu0}
\end{equation}
This equation was not used in the mentioned papers.

Finally, the definition of the observable ratio of densities is the following:
\begin{align}
\cfrac{\Omega_{DM}}{\Omega_b}
    &\begin{aligned}
        \approx\cfrac{3m_{UU}}{2m_p}\left|\cfrac{TB}{B}\right|+\cfrac{3m_{E/N}}{m_p}\left|\cfrac{TL}{B}\right|,
        \label{plotn}
    \end{aligned}
\end{align}
where it was assumed that the baryonic matter consists of protons only.

\section{Baryon and dark matter excesses}
\subsection{Before the electroweak phase transition}\label{s:before}
\subsubsection{Main dependencies}

To solve the system of equations which was described in previous section, a number of simplifying assumptions should be introduced. The first one has already been done in equation (\ref{plotn}): the baryonic matter consists of protons only. The second one is an assumption of equal masses: $m_U=m_D$ and $m_N=m_E$. By so, one can find
\begin{equation}
    \cfrac{TB}{B}=-\cfrac{\sigma_{UU}(3y\sigma_E-1)}{3y(\sigma_{UU}+3\sigma_E)}\left(\cfrac{L}{B}+\cfrac{9y\sigma_E+1}{3y\sigma_E-1}\right),
    \label{TBtoB}
\end{equation}
\begin{equation}
    \cfrac{TL}{B}=-\cfrac{\sigma_{E}(y\sigma_{UU}+1)}{y(\sigma_{UU}+3\sigma_E)}\left(\cfrac{L}{B}+\cfrac{3y\sigma_{UU}-1}{y\sigma_{UU}+1}\right).
    \label{TLtoB}
\end{equation}

\begin{figure}[t]
    \begin{minipage}{0.49\linewidth}
    \centering{\includegraphics[width = 1\linewidth]{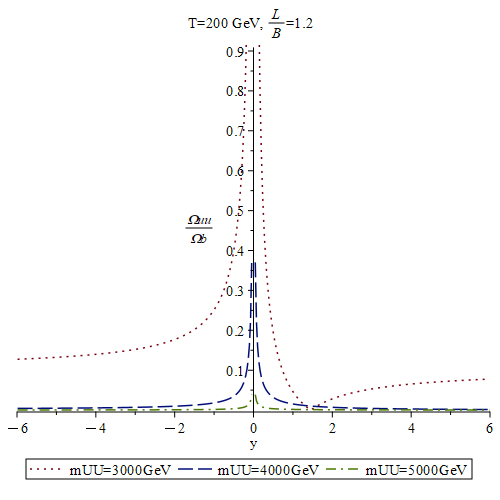}} 
    \end{minipage}
    \begin{minipage}{0.49\linewidth}
    \centering{\includegraphics[width = 1\linewidth]{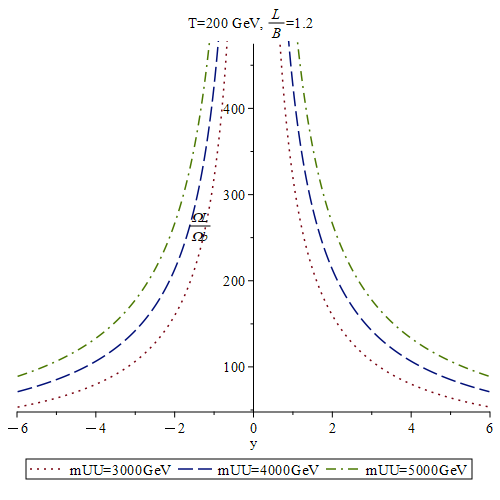}} 
    \end{minipage}
    \caption{Dependence of technibaryon (left) and technilepton (right) densities on the charge parameter $y$.}    
    \label{fig:Omy}
\end{figure}

The question on the relationship between the total masses of techniquarks (or technibaryons) and technileptons is more complicated. The simplest way is to set $m_{E/N}=\cfrac{m_{UU}}{2}$. In this case, the technibaryon component of the DM is exponentially suppressed by factor
\begin{equation}
    \cfrac{\sigma_{UU}(3y\sigma_E-1)}{3y(\sigma_{UU}+3\sigma_E)} \sim \cfrac{\sigma_{UU}}{\sigma_E}
\end{equation}
at high values of total mass. It also can be seen at Fig.\ref{fig:Omy}. One can see that the ratio of densities $\cfrac{\Omega_{DM}}{\Omega_b}$ depends on the electric charge parameter $y$ hyperbolically but the value of this ratio for the DM components may differ by several orders of magnitude.

\begin{figure}[t]
    \begin{minipage}{1\linewidth}
    \centering{\includegraphics[width = 1\linewidth]{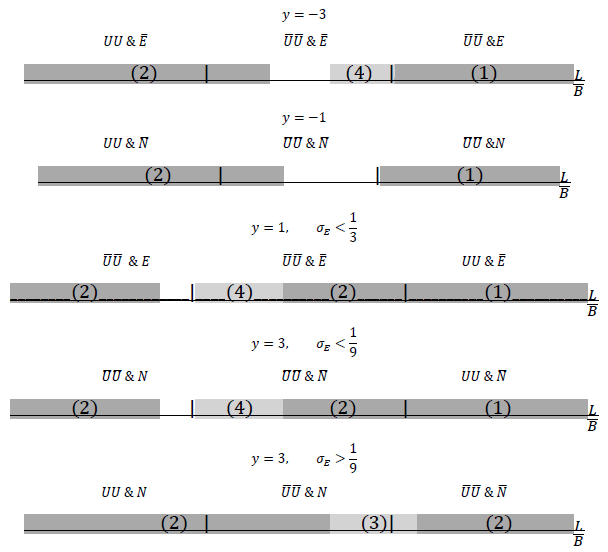}} 
    \end{minipage}
    \caption{The allowed regions of parameters for different values of charge.}    
    \label{fig:adep}
\end{figure}

The line break on the left panel of Fig. \ref{fig:Omy} describes the change of the sign of charges in the technibaryons excess (the excess of particles is replaced by an excess of antiparticles). This sign depends on many factors and has a strong impact on the observed physical picture which is shown on Fig.\ref{fig:adep}. The vertical lines points zero values of (\ref{TBtoB}) and (\ref{TLtoB}).

The grey areas on the picture are forbidden because
\begin{itemize}
    \item[(1)] both the technibaryons and the technileptons have a positive electric charge $+2n$. It leads to the overproduction of anomalous isotopes;
    \item[(2)] the overproduction of the technileptons: $\frac{\Omega_{E/N}}{\Omega_b} \gg 5 $;
    \item[(3)] the overproduction of the technibaryons: $\frac{\Omega_{UU}}{\Omega_b} \gg 5 $;
    \item[(4)] only the dominant DM component (technileptons) has a positive electric charge, it leads to the overproduction of anomalous isotopes too.
\end{itemize}
It should be noted, in order to avoid overproduction of the DM, it is necessary to consider the values of $\cfrac{L}{B}\sim1$.

The nature of the DM in allowed regions depends on the sign of the charge parameter $y$. If $y>0$, the DM should consist of both the technileptonic $(N/E)^{\frac{-3y\pm1}{2}}(^4He^{+2})_{\frac{-3y\pm1}{4}}$ and the technibaryonic $(\bar{U}\bar{U})^{y+1}(^4He^{+2})_{\frac{y+1}{2}}$ dark atoms in the X-helium form. But if it has a negative value, the WIMP like bound state $(\bar{U}\bar{U})_m(\bar{N})_n$ arise. Besides, the X-helium technileptonic, $\bar{N}^{-2r}$, and mixed, $\bar{N}^{-2r}(\bar{U}\bar{U})^{+2s}$, cores are possible in this case. 

It should be noted also that the case $y=-1$ is specific: the stable  technibaryon $UU$ has zero electric charge. So, it becomes the WIMP and the technileptonic X-helium $\bar{E}He$ is a dominant component of the DM.

\begin{figure}[t]
    \begin{minipage}{0.32\linewidth}
    \centering{\includegraphics[width = 1\linewidth]{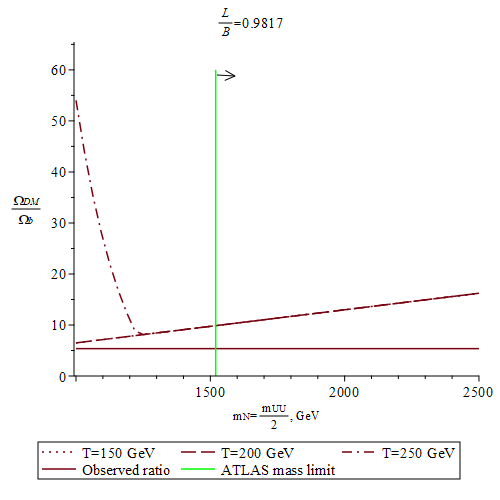}} 
    \end{minipage}
    \begin{minipage}{0.32\linewidth}
    \centering{\includegraphics[width = 1\linewidth]{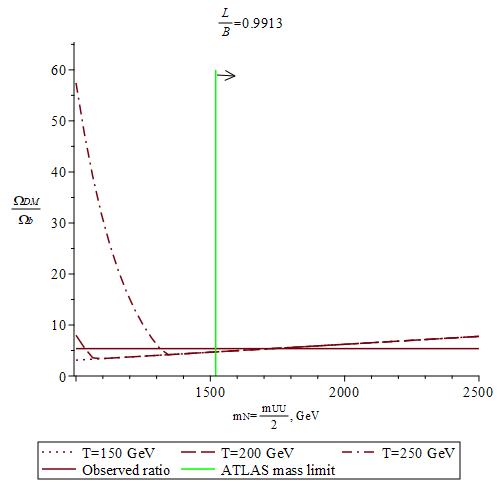}} 
    \end{minipage}
    \begin{minipage}{0.32\linewidth}
    \centering{\includegraphics[width = 1\linewidth]{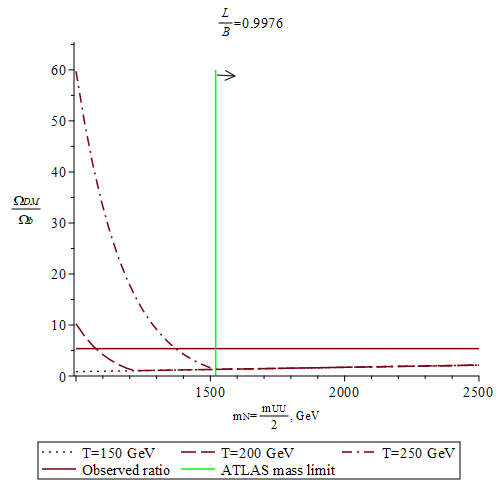}} 
    \end{minipage}
    \caption{The ratio of densities as the function of the total mass.}   
    \label{fig:Omm}
\end{figure}

On Fig.\ref{fig:Omm} typical behavior of the densities ratio as a function of the total mass is shown. It strongly depends on the value of ratio $\cfrac{L}{B}$. The ATLAS limits for the multi-charged particles masses can be seen in\cite{atlas}.

Using the information shown in Figs.\ref{fig:adep} and \ref{fig:Omm}, it can be found that the total mass of the techniparticles can not be lighter than $1\,\mbox{TeV}$. Unfortunately, there are no upper limits for these masses in general case, but in special cases the possibility of the DM overproduction allows to set some restrictions. 

\subsubsection{Different masses}

\begin{figure}[t]
    \begin{minipage}{0.32\linewidth}
    \centering{\includegraphics[width = 1\linewidth]{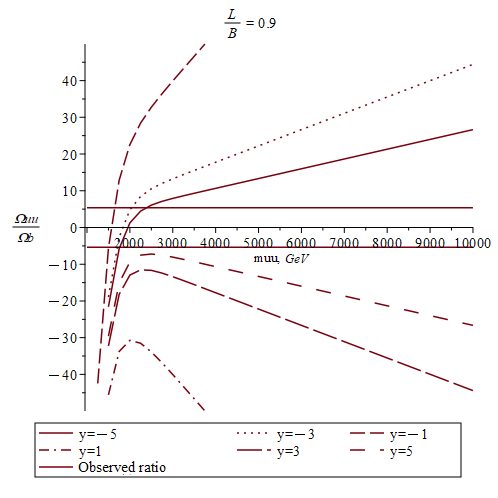}} 
    \end{minipage}
    \begin{minipage}{0.32\linewidth}
    \centering{\includegraphics[width = 1\linewidth]{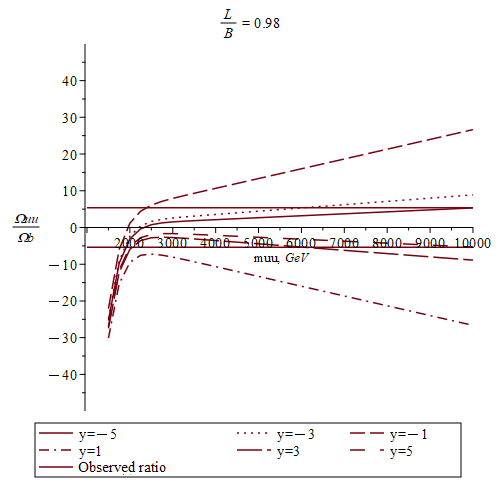}} 
    \end{minipage}
    \begin{minipage}{0.32\linewidth}
    \centering{\includegraphics[width = 1\linewidth]{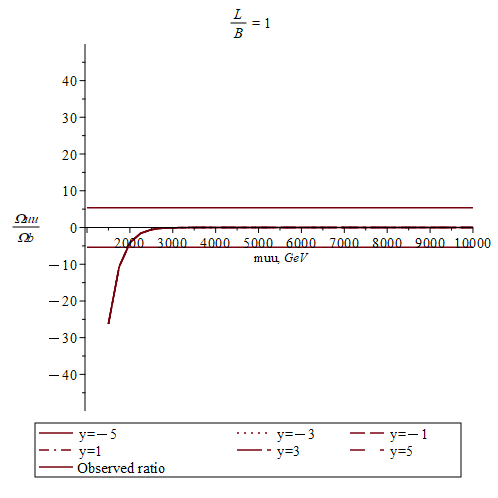}} 
    \end{minipage}
    \\
    \begin{center}
    \begin{minipage}{0.32\linewidth}
    \centering{\includegraphics[width = 1\linewidth]{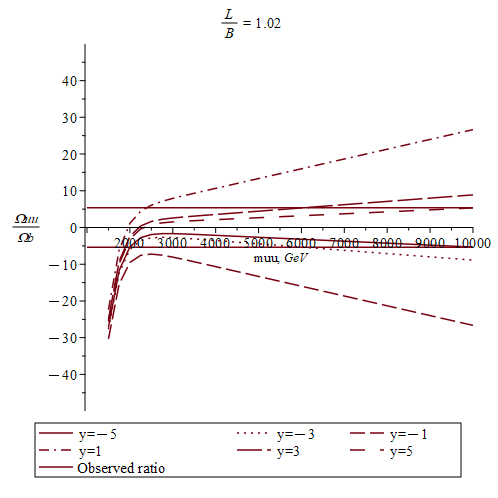}} 
    \end{minipage}
    \begin{minipage}{0.32\linewidth}
    \centering{\includegraphics[width = 1\linewidth]{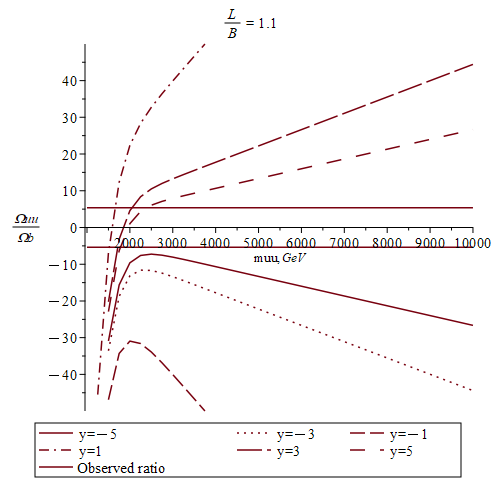}} 
    \end{minipage}
    \end{center}
    \caption{The ratio of densities for technibaryon component of the DM (assuming $m_{E/N}={m_{UU}}$) as a function of total mass.}    
    \label{fig:assump}
\end{figure}

It is possible to study how the mass difference changes the described picture. First of all,assumption on the mass values should be considered more carefully. The equation $m_{E/N}=\cfrac{m_{UU}}{2}$ which has been used above could be replaced by $m_{E/N}={m_{UU}}$. Then, there is no suppression of technibaryonic component of the DM. It is shown on Fig.\ref{fig:assump} how the density of technibaryons depends on the total mass under the new assumption.The allowed parameters regions on Fig.\ref{fig:adep} in this case should decrease.

\begin{figure}[t]
    \begin{minipage}{0.32\linewidth}
    \centering{\includegraphics[width = 1\linewidth]{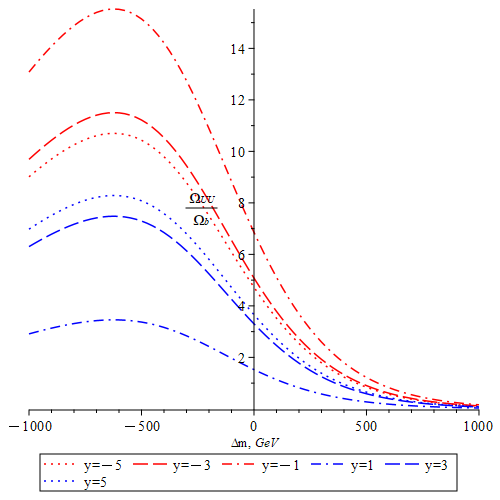}} 
    \end{minipage}
    \begin{minipage}{0.32\linewidth}
    \centering{\includegraphics[width = 1\linewidth]{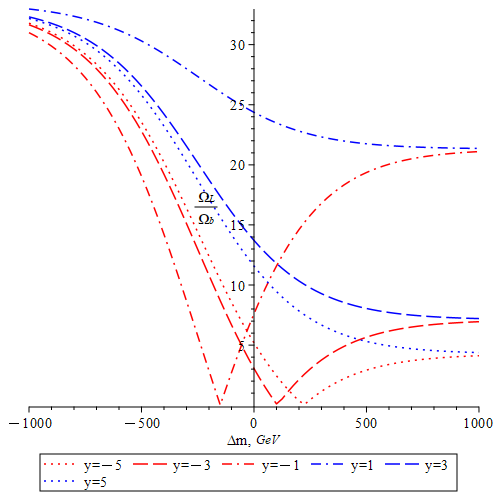}} 
    \end{minipage}
    \begin{minipage}{0.32\linewidth}
    \centering{\includegraphics[width = 1\linewidth]{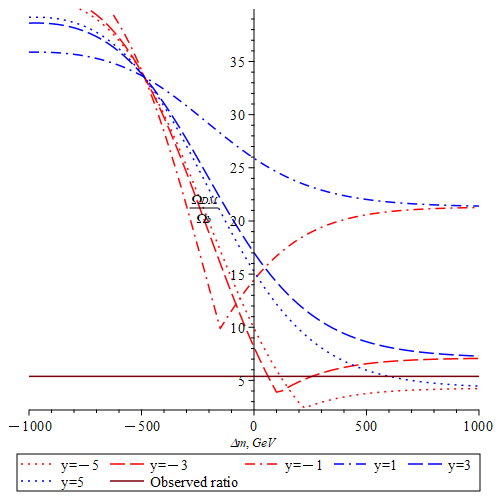}} 
    \end{minipage}
    \caption{The general dependence of ratios of densities on the mass difference $\Delta m = m_{UU}-m_{E}$.}    
    \label{fig:dm}
\end{figure}

In general, due to the mass difference, $\Delta m = m_{UU}-m_{E}$, the sign of the generated excess can change. It is shown in Fig.\ref{fig:dm} for the density of technibaryon component (left panel), density of technilepton component (central panel) and total density of the DM (right panel).

To take into account the difference between the masses of technibaryons and technileptons, it is necessary to enter the following parametrization:
\begin{align}
    &\sigma_{N,E} = \sigma_f\left(\cfrac{m}{T}\right)+n,e;
    \\
    &\sigma_{UU}=\sigma_b\left(\cfrac{m}{T}\right)+2u;
    \\
    &\sigma_{UD}=\sigma_b\left(\cfrac{m}{T}\right)+u+d;
    \\
    &\sigma_{DD}=\sigma_b\left(\cfrac{m}{T}\right)+2d,
\end{align}
where $m=1500\, \mbox{GeV}$ and $T=250\, \mbox{GeV}$ are chosen to minimize the physically meaningful value of the weight function.

\begin{figure}[t]
    \begin{minipage}{0.24\linewidth}
    \centering{\includegraphics[width = 1\linewidth]{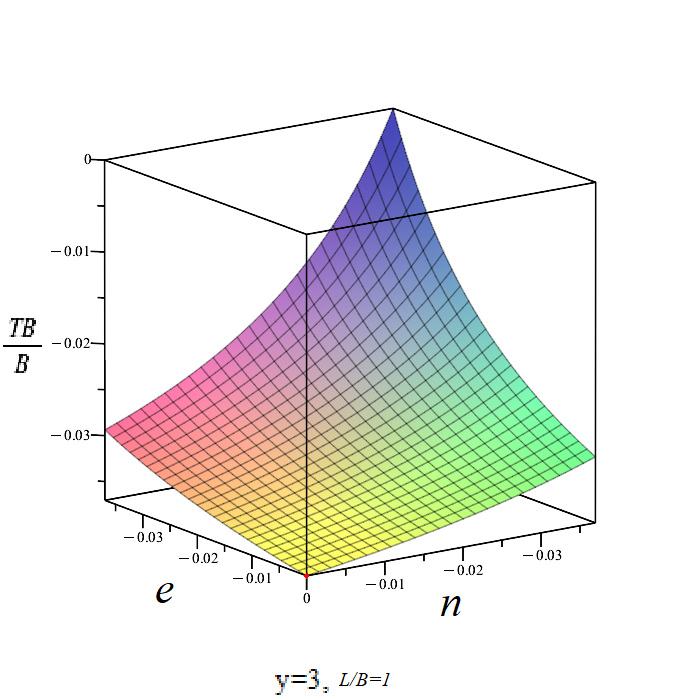}} 
    \end{minipage}
    \begin{minipage}{0.24\linewidth}
    \centering{\includegraphics[width = 1\linewidth]{Fig/Before/dif_m/WTC.Bef.eq3.a1TB.ne.png}} 
    \end{minipage}
    \begin{minipage}{0.24\linewidth}
    \centering{\includegraphics[width = 1\linewidth]{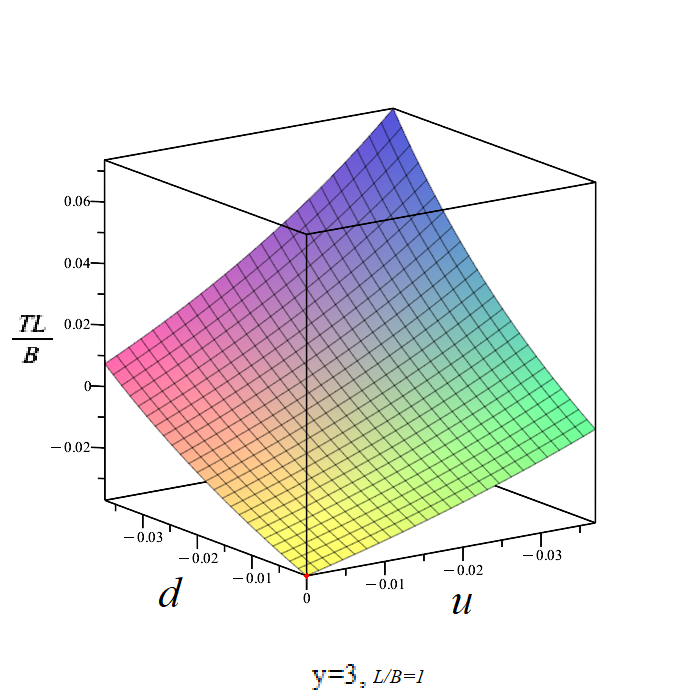}} 
    \end{minipage}
    \begin{minipage}{0.24\linewidth}
    \centering{\includegraphics[width = 1\linewidth]{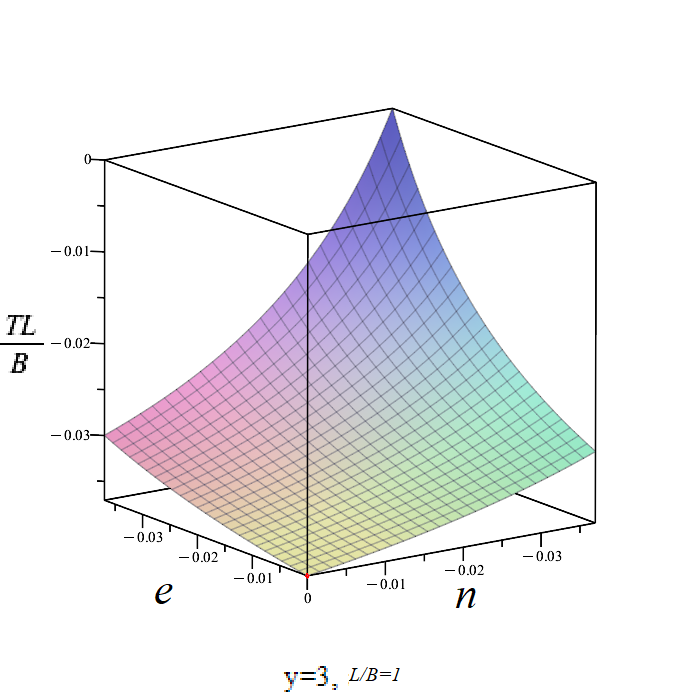}} 
    \end{minipage}
    \\
    \begin{minipage}{0.24\linewidth}
    \centering{\includegraphics[width = 1\linewidth]{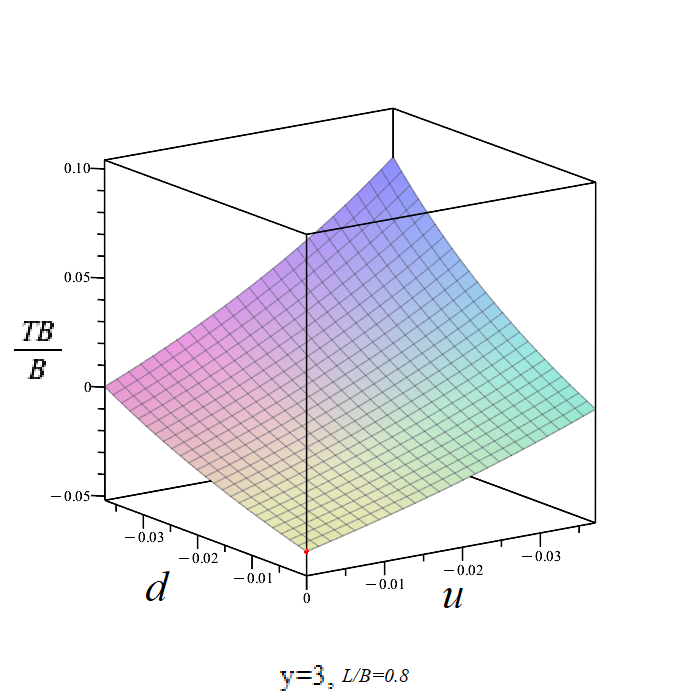}} 
    \end{minipage}
    \begin{minipage}{0.24\linewidth}
    \centering{\includegraphics[width = 1\linewidth]{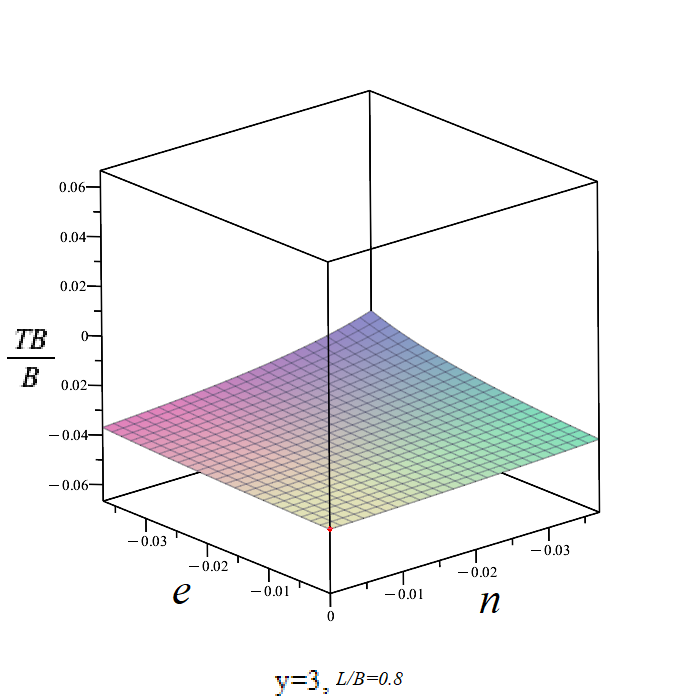}} 
    \end{minipage}
    \begin{minipage}{0.24\linewidth}
    \centering{\includegraphics[width = 1\linewidth]{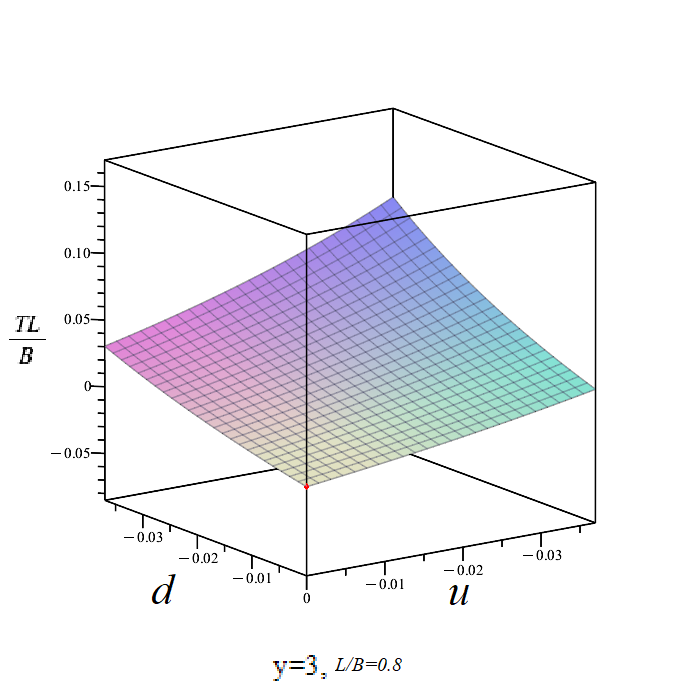}} 
    \end{minipage}
    \begin{minipage}{0.24\linewidth}
    \centering{\includegraphics[width = 1\linewidth]{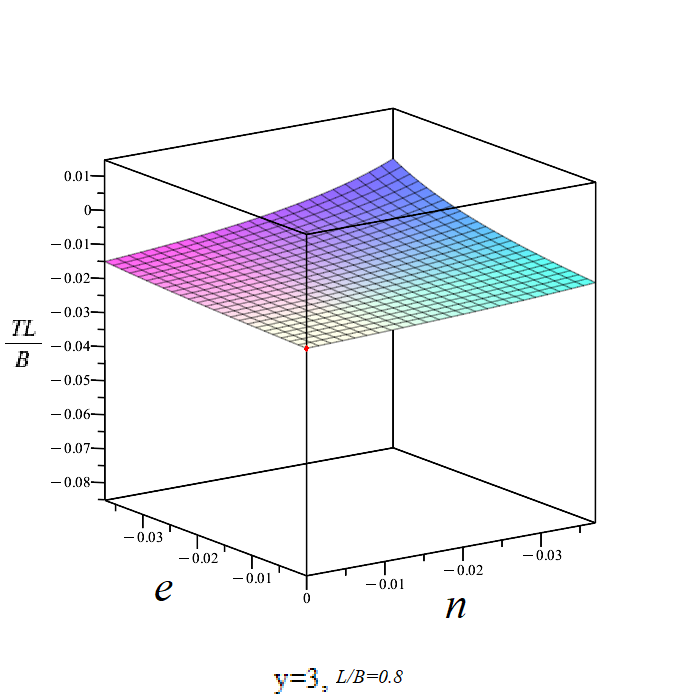}} 
    \end{minipage}
    \caption{The ratios of densities for different techniparticle masses.}
    \label{3dmass}
\end{figure}

One can see in Fig.\ref{3dmass} that the mass difference increasing results to rise of the density ratio for both DM components. Moreover, in some cases the sign of the generated techniparticles excess can be changed.

\subsection{After the electroweak phase transition}\label{s:after}

For temperatures below the EWPT temperature, condition of isospin neutrality can not be used, so the system of equations solution is:
\begin{equation}
    \cfrac{TB}{B}=-\alpha~\left(\cfrac{L}{B}+\gamma~\cfrac{TL}{B}+\beta\right),
    \label{pTBtoB}
\end{equation}
where the functions $\alpha$, $\beta$ and $\gamma$ are considered assuming an equal masses: 
\begin{align}
    &\alpha=\cfrac{\sigma_{UU}}{3}~\cfrac{(\sigma_{t}+5)(2\sigma_{UU}+\sigma_{E})+6(\sigma_{t}+17)}{( 9(\sigma_{t}-1)y+2(\sigma_{t}+5) )\sigma_{UU}+(\sigma_{t}+5)\sigma_{E}+3(5\sigma_{t}+31)},
    \label{alpha}
    \\
&\beta=\cfrac{18(2\sigma_{UU}+\sigma_{E}+18)}{(\sigma_{t}+5)(2\sigma_{UU}+\sigma_{E})+6(\sigma_{t}+17)},
\label{beta}
    \\
    &\gamma=\cfrac{2(\sigma_{t}+5)\sigma_{UU}+( 27(1-\sigma_{t})y+\sigma_{t}+5 )\sigma_{E}+3(5\sigma_{t}+31)}{\sigma_{E}((\sigma_{t}+5)(2\sigma_{UU}+\sigma_{E})+6(\sigma_{t}+17))}.
    \label{gamma}
\end{align}

For $|y|\lesssim100$, dependence of the density ratio on the charge can be neglected. So, the result of calculations depends only on two parameters: $\cfrac{L}{B}$ and $\cfrac{m}{T_*}$.

\begin{figure}[t]
    \begin{center}
    \begin{minipage}{0.32\linewidth}
    \centering{\includegraphics[width = 1\linewidth]{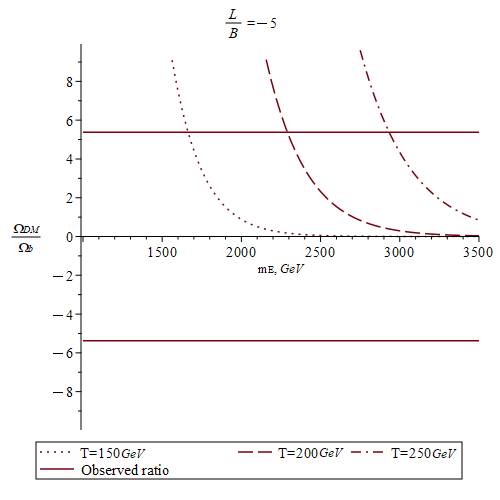}} 
    \end{minipage}
    \begin{minipage}{0.32\linewidth}
    \centering{\includegraphics[width = 1\linewidth]{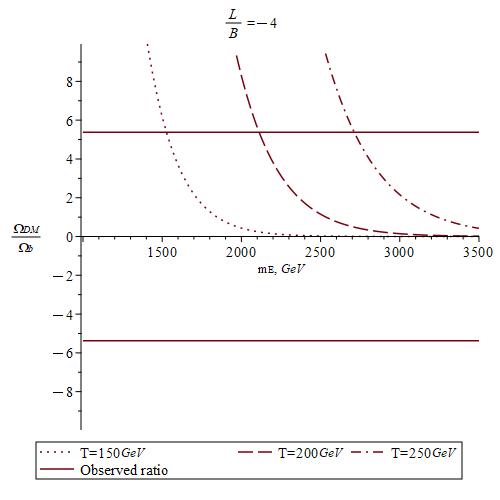}} 
    \end{minipage}
    \end{center}
%    \\
    \begin{minipage}{0.32\linewidth}
    \centering{\includegraphics[width = 1\linewidth]{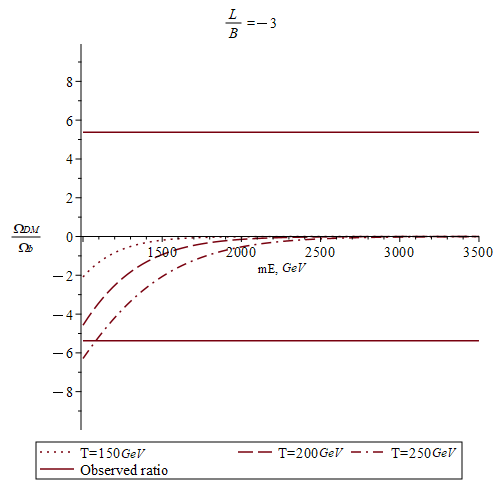}} 
    \end{minipage}
    \begin{minipage}{0.32\linewidth}
    \centering{\includegraphics[width = 1\linewidth]{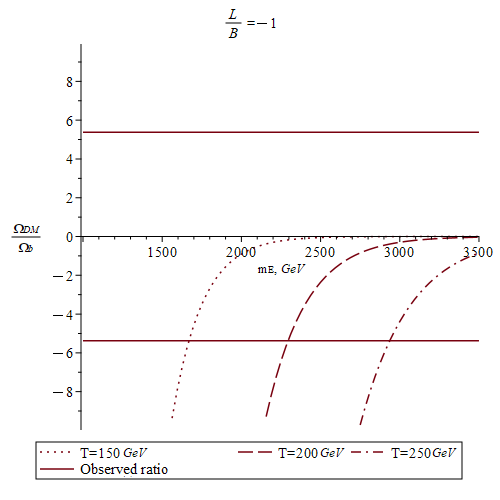}} 
    \end{minipage}
    \begin{minipage}{0.32\linewidth}
    \centering{\includegraphics[width = 1\linewidth]{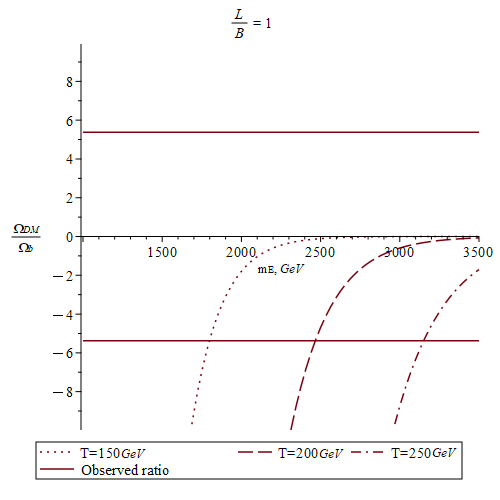}} 
    \end{minipage}
    \caption{The ratio of densities as a function of two main parameters for $T_*<T_{EWPT}$.}    
    \label{fig:After}
\end{figure}

If $\Delta m= m_{UU}-m_{E} \not=0$ the technibaryon component of the DM should been suppressed by factor  
\begin{equation}
    \alpha \gamma \xrightarrow{\sigma_{t}\rightarrow1} \cfrac{1}{3}~\cfrac{\sigma_{UU}}{\sigma_{E}}.
\end{equation}
It allows to find the value of densities ratio for the technileptonic DM assuming $\cfrac{\Omega_{UU}}{\Omega_b}\rightarrow0$. In Fig.\ref{fig:After} one can see how the ratio depends on two main parameters; negative values corresponds to the excess of antitechnileptons.

As  it is seen, the expected mass of techniparticles becomes higher with the increasing of $\cfrac{L}{B}$ parameter. Because $\cfrac{L}{B}<10^8$ (see\cite{LtoB}), the upper limit for the technileptonic mass can be estimated as: $m<5-8\, \mbox{TeV}$ for different sphaleron freezing out temperatures.

\section{Conclusion}\label{s:concl}

The cosmological consequences of sphaleron transitions in the  minimal walking technicolor model have been considered. The main feature of  the model is the nonfixed arbitrary electric charge of new heavy particles. Different values of the charge parameter $y$ lead to different scenarios of multicomponent DM.

It was shown, the ratio of densities $\frac{\Omega_{DM}}{\Omega_b}$ could be explained in both considered cases ($T_*>T_{EWPT}$ and $T_*<T_{EWPT}$). If sphaleron transitions freeze out before the EWPT, the DM should consist of several forms of X-helium and/or WIMP-like bound states. To prevent overproduction of particles, it is necessary to set a lower limit on their mass  $m\gtrsim1\,\mbox{TeV}$. The upper limit can be found only for some special cases.

If sphaleron transitions freeze out after the EWPT and difference  between the masses of technibaryons and technileptons can not be neglected, the DM particle should not be too heavy ($m<5-8\, \mbox{TeV}$). In this case the DM density is provided by the X-helium with technileptonic core.

 The found limits for masses could be tested in upcoming accelerator experiments.
\section*{Acknowledgements}

The research by V.B. and M.K. was carried out in Southern Federal University with financial support of the Ministry of Science and Higher Education of the Russian Federation (State contract GZ0110/23-10-IF). The research by D.S. was performed in the framework of MEPHI program on Prioritet 2030.

%% The bibliography section

\end{document}